\documentstyle[psfig,conf_iap,10pt]{article}
\begin{document}
\newcommand{\apg}{\:^{>}_{\sim}\:}
\newcommand{\apl}{\:^{<}_{\sim}\:}
\newcommand{\cmjj}{\mbox{${\rm cm^{-2}}$}}
\newcommand{\etal}{et al.}
\newcommand{\hI}{\mbox{${\rm H\ I}$}}
\newcommand{\htwo}{\mbox{${\rm H_2}$}}
\newcommand{\ibid}{\underline{\makebox[0.5in]{}}.}
\newcommand{\iue}{IUE}
\newcommand{\hst}{HST}
\newcommand{\kms}{\mbox{km\ s${^{-1}}$}}
\newcommand{\lya}{\mbox{${\rm Ly}\alpha$}}
\newcommand{\lyb}{\mbox{${\rm Ly}\beta$}}
\newcommand{\lyc}{\mbox{${\rm Ly}\gamma$}}
\heading{%
A High Deuterium Abundance at z=0.7; Evidence for 
Cosmic Inhomogeneity?
}
\par\medskip\noindent
\author{J. K. Webb$^1$, R. F. Carswell$^2$, K. M. Lanzetta$^3$, 
R. Ferlet$^4$, M. Lemoine$^5$, A. Vidal-Madjar$^4$}
\address{School of Physics, University of New South Wales, Sydney, 
NSW 2052, Australia}
\address{Institute of Astronomy, Madingley Road, Cambridge, CB3
0HA, England}
\address{Department of Physics and Astronomy, State University of 
New York at Stony Brook, NY 11794-3800, USA}
\address{Institut d'Astrophysique de Paris, CNRS, 98bis Boulevard Arago,
F--75014 Paris, France}
\address{Department of Astronomy \& Astrophysics, Enrico Fermi
Institute, University of Chicago, Chicago IL 60637--1433, USA}

\begin{abstract}
Recent HST/GHRS observations of the $z = 0.7010$ absorber toward the 
QSO 1718$+$4807 (selected because of apparently ideal characteristics
for measuring D/H) yield a D/H significantly higher D/H than some recent
high-redshift measurements.  
Our analysis indicates ${\rm D/H} = 1.8 - 2.5 \times 10^{-4}$.  
This may indicate a cosmological inhomogeneity in the deuterium abundance 
of at least a factor of ten.
\end{abstract}
\section{Introduction, target selection and observations}
Measuring D/H in QSO absorbers is
observationally difficult for a number of reasons: (1) ill-placed H~I
\lya-forest absorption lines can masquerade as deuterium and (2)
most QSO absorption systems exhibit complex internal velocity
structure, which may render parameter estimation for individual
components of interest unreliable or impossible.   
Here we report results from one object which
appears to have the ideal characteristics for a D/H analysis.
Only one absorbing component is revealed by the data and the
velocity dispersion in that component is small enough to easily detect
and accurately measure the deuterium abundance.
 
The absorber at redshift $z = 0.7010$ toward the QSO 1718$+$4807
($z_{\rm em} = 1.084$, $m_V = 15.3$) was selected (by KML) on the basis of a
remarkably abrupt partial Lyman-limit discontinuity in the International
Ultraviolet Explorer (IUE) spectrum \cite{lts93}.  The extreme sharpness of the
Lyman break clearly indicates simple velocity structure {\it and} low
velocity dispersion parameter.  HST observations were obtained in Cycle 4
using the Goddard High Resolution Spectrograph (HRS) and the G270M
grating of the spectral region covering the \lya\ and Si~III absorption
lines.  The spectra were extracted, binned to linear wavelength
scales and corrected to vacuum, heliocentric wavelengths using
standard procedures.

\section{Parameter estimation}
Parameter estimates were derived using VPFIT \cite{w87},
an unconstrained optimisation $\chi^2$ minimisation 
(Gauss-Newton) algorithm.  
Descent to the minimum in $\chi^2$-parameter space is obtained 
using first and second derivatives of $\chi^2$ with respect to the 
free parameters.  Reliable parameter error estimates are obtained
from the parameter covariance matrix.

When simultaneously fitting several lines in the same QSO 
absorption system, we are able to `tie' physically related 
parameters, resulting in a more
stringent test of the model being fitted, and tighter constraints on the
remaining free parameters.  The simplest example is
simultaneously fitting a single cloud with D and H.  6 parameters are
required to fit as two individual features, but only 4 
($z$, $N({\rm HI})$, $N({\rm DI})$, $b({\rm HI}) $) if physically
related parameters are tied; we can constrain the Doppler parameter 
for deuterium $b$(DI) to lie between $b$(HI) and $b({\rm HI})/\sqrt{2}$.

Combining several different datsets of very different spectral
resolution and signal-to-noise (as is the case here where
IUE and HST data are analysed simultaneously) is straightforward, 
minimising a
single $\chi^2$, where contributions to the overall $\chi^2$
from individual datasets are appropriately weighted.
\section{D/H for single cloud model}
\subsection{Fits to \lya\, SiIII, Lyman limit}
Although the cloud parameters are determined in the manner above, i.e.
by simultaneously fitting all parameters to both data sets, the
parameter constraints arise essentially as follows.  The IUE spectrum
of the partial Lyman limit provides an extremely accurate
determination of the H~I column density ($\log N({\rm H~I}) = 17.24 \pm
0.01$ \cmjj).  The Si~III $\lambda$1206 absorption line supports the single
component velocity structure implied by the IUE Lyman limit and
determines the cloud redshift precisely ($z_{\rm abs} = 0.701024 \pm
0.000007$).  The \lya\ absorption is clearly asymmetric with respect
to the Si~III redshift, showing additional absorption in the blue wing
at the position corresponding to D~I (Figure 1).  Since Si~III
accurately constrains the position of the D~I feature and because the
Doppler parameter $b({\rm D~I})$ is constrained by the overall system fit,
only one free parameter, $N({\rm D~I})$, is required to fit the excess
absorption seen at the position corresponding to D.  The D~I column
density is thus accurately determined and is $\log N({\rm D~I}) = 13.57 \pm
0.06$ \cmjj.  The Doppler parameters are dominated by non-thermal broadening
with an inferred cloud temperature of $1.9 \times 10^4$ K and Doppler
parameters of $b({\rm H~I}) =  25.5$ \kms\ and $b({\rm D~I}) = 22.2$ \kms\
and $b({\rm Si~III}) = 18.7$ \kms, with the same error on each of 
$\sigma = 0.5$ \kms.  From the above, the measured deuterium abundance lies 
in the range ${\rm D/H} = 1.8 - 2.5 \times 10^{-4}$.
\begin{figure}
\centerline{\vbox{
\psfig{figure=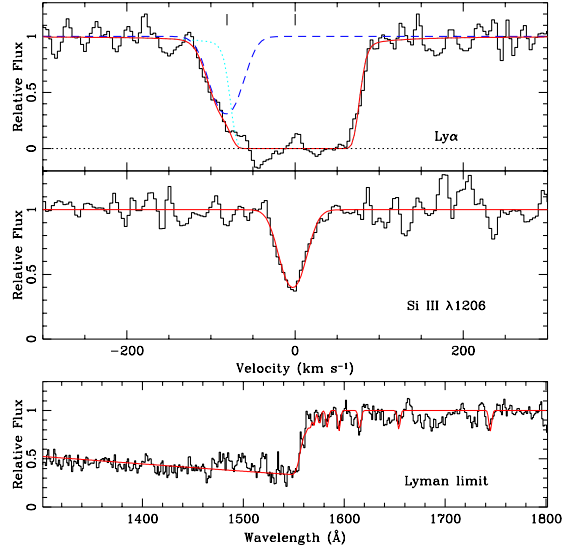,height=8.cm}
}}
\caption[]{Best fit to GHRS \lya\, Si~III, and IUE Lyman limit
}
\end{figure}
\subsection{Fits to \lya\ and Lyman limit only, excluding Si~III}
The strongest Si~III component in an
absorption complex need not align with the strongest H~I
component.  We thus investigate whether the results above
are sensitive to the assumption that $z({\rm Si~III}) = z(\lya)$
by fitting \lya\ and the Lyman limit only.
The H~I redshift is no longer constrained by that of Si~III.
A priori one expects a larger error on the remaining column
density estimates, plus the D~I column density could change if Si~III
is shifted with respect to H~I.  The result of doing this is that
$\log N({\rm D~I}) = 13.62 \pm 0.06$ \cmjj, so 
the measured deuterium abundance lies 
in the range ${\rm D/H} = 1.8 - 3.1 \times 10^{-4}$.
\subsection{Why D/H is unlikely to be low}
We can {\it fix} ${\rm D/H} = 2.5 \times 10^{-5}$, re-fit allowing all
other parameters to vary, and compare the resulting $\chi^2_{min}$
with the best fit $\chi^2_{best}$ derived with both H~I and D~I column
densities free to vary.  The result is a substantially worse fit, with
$\chi^2_{min} = \chi^2_{best} + 57$.  The best fit with D/H
constrained to be low is illustrated in Figure 2.  Note the poor
fit to the blue wing of the \lya~ feature and
how the system redshift has been pushed slightly lower in order to
partially compensate for the additional absorption at the D~I
position (especially noticable in Si~III).  In the next section
we discuss why D/H is unlikely to be low, even in the
presence of a second badly blended absorbing component.
\begin{figure}
\centerline{\vbox{
\psfig{figure=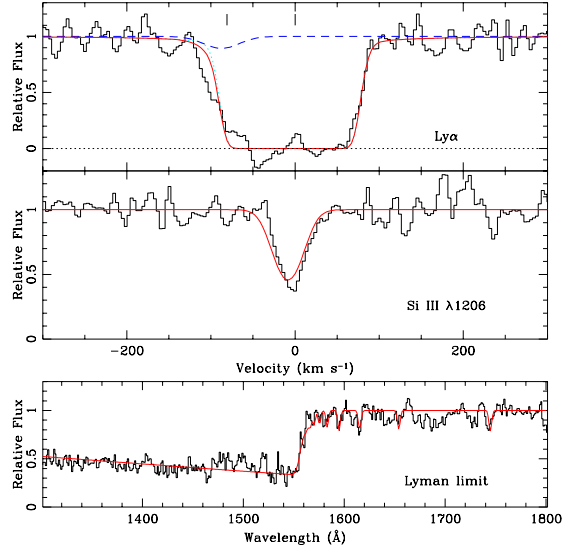,height=8.cm}
}}
\caption[]{Best fit to GHRS \lya\, Si~III, and IUE Lyman limit, with D/H 
constrained to be $2.5 \times 10^{-5}$
}
\end{figure}
\subsection{Why a single cloud model is most appropriate}
The simple velocity structure here contrasts with the more
complex nature of the higher redshift systems reported so far,
both those which give low D/H measurements \cite{bt96, tfb96, tbk96} 
and those which give high values \cite{c94, rh96a, rh96b, swc97}.  
To examine 
the possibility that a weak H~I cloud is present somewhere near the
expected position of D~I, we re-fitted the data replacing D~I with
H~I.  The best fit resulted in the new H~I line falling at $-86 \pm 5$
\kms\ from the strong component, which agrees with the expectation for
D~I in both direction and magnitude.  Furthermore, the Doppler
parameter of this putative H~I interloper ($b = 21 \pm 4$ \kms) lies
between that of the strong H~I component but greater than the Si~III
value, which is again consistent with the expectation for D I.  
Finally, if we assume a {\it random} distribution of clouds, the
observed line number density suggests a probability of an interloper
falling within $\pm 4\sigma$ of the actual position of D~I of
$\sim 1$\%.  Whilst a random distribution will not generally
represent the underlying cloud distribution in any complex, we
note that this particular absorber was selected specifically on
the basis of a sharp Lyman limit.  It would therefore be
incorrect to estimate the interloper probability on the basis of
a cloud-cloud correlation function derived from a larger sample of 
absorbers, where {\it no} such selection had taken place.

We can explore the possibility of multiple structure further,
by artificially inserting additional components and re-fitting.
We have done this in two independent ways.  First, we
pursue the possibility above, i.e. a potentially low D/H,
but this time assuming velocity structure is present in the cloud.
To do this we model the spectrum with a double cloud where
D/H is forced to be the same in each, and re-fit
for a range in velocity separation from 5 to 81 \kms.  The
upper value corresponds to the splitting between D~I and H~I.
For no value of the double cloud separation do we find a value 
of $\chi^2$ {\it below} that for the single cloud.  Furthermore,
for a separation of less than 30 \kms, the $\chi^2$ lies
at over $3 \sigma$ {\it above} the single cloud value.
Thus a double cloud model appears unable to mimic a high D/H.

To see what the inferred  D/H is in the presence of a second
cloud, we take an illustrative double cloud separation of
10 \kms, and re-fit allowing all other parameter to vary.
In this case, a careful choice of free parameters is essential;
the Lyman limit constrains the {\it total} N(H~I) extremely well,
although the two individual N(H~I) are very poorly determined.
For this reason, the H~I column density parameters are the
{\it total} N(H~I) and N(H~I) for one of the clouds.  Although we
maintain the same number of free parameters, this choice results
in well determined N(H~I) for each cloud.  We can also choose
whether to force D/H to be the same in each cloud, or allow
it to vary.
The range of results deduced from both options is
${\rm D/H} = 1.3 - 3.7 \times 10^{-4}$, so the earlier
general conclusion remain unaltered.

Taken together, these points indicate that the most reasonable and 
likely interpretation of the data is that we have detected D~I.

\section{Discussion}

\subsection{A Comment on the reliability of low D/H values}

A hotly debated issue during this session concerns the reliability
of the low D/H measured towards 1937-1009 \cite{tfb96, swc97, wamp96},
the main points being that the inferred D/H could be somewhat
larger if N(H~I) has been underestimated, either due to errors
in establishing the continuum level below the Lyman limit, or
if undetected velocity structure is present.  Two methods
have been discussed addressing the former point: (a) a statistical approach
(modeling the statistical \lya\ distribution) and (b) a `spectrum-specific'
approach (estimating N(H~I)'s for each and every cloud in the spectrum 
which is able to contribute opacity to the relevant regions of the 
Lyman limit).  We merely comment here that although method (a) can in
principle estimate the {\it average} opacity to arbitrarily high
accuracy, fluctuations from one sightline to the next are likely
to be large.  Method (b) will also be error-prone because
for any given ensemble of absorption lines, {\it some} clouds will
have indeterminable parameters.  The most believable approach is
to treat the continuum level below the Lyman limit as an additional
free parameter.

However, we comment finally that whilst there may be some uncertainty
in the value of D/H estimated towards 1937-1009, it is unlikely to
be great enough to change D/H to be as high as the value we derive for
1718+4807.

\subsection{Cosmic D/H inhomogeneity?}

A ratio ${\rm D}/{\rm H}=2.0 \times 10^{-4}$, as obtained here toward
1718+4807, corresponds to a mass fraction of $^4$He $Y \simeq 0.233
\pm 0.002$, $^7{\rm Li}/{\rm H} \simeq 1.8^{+0.7}_{-0.6} \times
10^{-10}$ and $\Omega_B \simeq 0.006 \pm 0.003 \ h^{-2}$, where $h$
denotes the value of the Hubble constant in units of 100 km s$^{-1}$
Mpc$^{-1}$ and the errors only include $1 \sigma$ errors on the
nuclear cross-sections of BBN.  Remarkably, these figures agree
precisely with the measured abundances, $Y \simeq 0.233 \pm 0.005$ and
$^7{\rm Li}/{\rm H} \simeq 1.8^{+0.5}_{-0.3} \times 10^{-10}$. The
above baryonic mass density parameter may be compared with the
measured amount of visible matter in galaxies, $\Omega_{\rm vis} \sim
0.002-0.005 \ h^{-1}$. This shows that a high primordial deuterium
abundance leaves little room for {\it baryonic} dark matter and
supports the view that the missing mass inferred from dynamical
studies must be non-baryonic.  On the other hand, a low $({\rm D}/
{\rm H})_p$ implies BBN values of $^4$He and $^7{\rm Li}/{\rm H}$
which differ from the measured values by amounts corresponding to $2
\sigma$ observational limits (although it removes any difficulty in
explaining rapid deuterium destruction).

The interpretation above sidesteps the apparently significant
difference between our measurement and the low value claimed in
reference \cite{tfb96}, where ${\rm D/H} = 2.3 \pm 0.6 \times
10^{-5}$, compared to ours of ${\rm D/H} = 1.8 - 2.5 \times 10^{-4}$
($1\sigma$) or ${\rm D/H} = 1.5 - 3.0 \times 10^{-4}$ ($2\sigma$).
Whilst further observations might correct one or both of these
measurements, the most reasonable interpretation of the data at the
moment is that the difference is real.  We speculate that these
observations may be the first to reveal the presence of fluctuations
in the baryon-to-photon number at the epoch of big bang
nucleosynthesis \cite{cf96, fc96}.
%



%
\vfill
\end{document}